\documentstyle[prl,aps,epsfig,amsfonts,amssymb,amsmath] {revtex}

\begin{document} 
\advance\textheight by 0.2in 
\draft
\twocolumn[\hsize\textwidth\columnwidth\hsize\csname@twocolumnfalse%
\endcsname
\title{Critical Behaviour of Non-Equilibrium Phase Transitions \\  to
Magnetically Ordered States} 
\author{Thomas Birner, Karen Lippert, Reinhard
M\"uller, Adolf K\"uhnel, and Ulrich Behn} 
\address{Institut f\"ur Theoretische
Physik, Universit\"at Leipzig, Augustusplatz 10, D-04109 Leipzig, Germany}
\date{\today} 
\maketitle %

\begin{abstract} We describe non-equilibrium phase transitions in arrays of
dynamical systems with cubic nonlinearity driven by
multiplicative Gaussian white noise. Depending on the sign of the  spatial
coupling we observe  transitions  to  ferromagnetic or antiferromagnetic
ordered states.  We discuss the phase diagram, the order of the
transitions, and the critical behaviour.  For  global coupling
we show analytically that the critical exponent of the
magnetization exhibits a transition from the value $1/2$ to a
non-universal behaviour depending on the ratio of noise strength to the
magnitude of the spatial coupling.  

\end{abstract}
%
] 

\noindent {\em Introduction.} In the last decade studying arrays of
stochastically driven nonlinear dynamical systems the notion of noise induced
non-equilibrium phase transition has been established
\cite{Shiino,Broeck94,Parrondo96,Garcia96,Broeck97,Grinstein96,Genovese98a,Genovese99,Mueller97,Kim97a};
for a recent monograph see \cite{GOS99}. In close analogy to equilibrium phase
transition one has order parameters and finds continuous or discontinuous
transitions associated with ergodicity breaking. The behaviour near the
transition point is characterized by power laws and a critical slowing down.

In this paper we consider arrays of spatially harmonically coupled Stratonovich
models \cite{Schenzle79} which undergo transitions into ordered states
comparable to ferromagnetic (FM)  or  antiferromagnetic (AFM) phases depending
on the sign of the coupling constant. The AFM situation is described  first in
this paper for that  class of models. We determine the phase diagram and
characterize the critical behaviour at these transitions. For the globally
coupled system we derive  an analytical result for the critical exponent of the
order parameter, i.e. the magnetization. This critical exponent exhibits a
hitherto not described  transition from a value $1/2$ to a non-universal
behaviour when increasing the ratio of noise strength and magnitude of the
spatial coupling.

The dynamics of the individual  constituents $x_i$ at the lattice sites
$i=1,\ldots,L $ is
governed by a system of stochastic ordinary differential equations in the
Stratonovich sense
\begin{equation}\label{langevin}
\dot{x}_i  = a x_i  -x_i^3  + x_i\, \xi_i  - \frac{D}{N} 
\sum_{j\in {\cal N}(i)} (x_i - x_j )\;,
\end{equation}
where ${\cal N}(i)$ denotes the set of sites interacting with site $i$. $N =
\# {\cal N}(i) $ is equal to $L-1$ in the case of global coupling and to $2d$
in the case of nearest neighbour (n.n.) coupling on a simple cubic lattice in
$d$ dimensions.  $D$ is the  strength of the spatial interactions.  $\xi_i(t)$
is a zero mean spatially uncorrelated Gaussian white noise with autocorrelation
function
$
\langle \xi_i(t)\, \xi_j(t') \rangle \, =  
\sigma^2 \delta_{i j} \delta(t -t'),
$
where $\sigma^2$ is the noise strength. 

The stationary probability density $P_s(x_i)$ fulfills the (reduced) 
stationary Fokker-Planck equation \cite{Broeck94}
\begin{eqnarray} \label{fokker}
0  = && \frac{\partial}{\partial x_i} \bigg [ \bigg (-a x_i+ x_i^3+
\frac{D}{N}\sum_{j \in {\cal N}(i)}(x_i-\langle x_j | x_i \rangle)\nonumber \\
    && +  \frac{\sigma^2}{2} x_i \frac{\partial}{\partial x_i} x_i \bigg )
P_s (x_i) \bigg ]\;, 
\end{eqnarray}
where $\langle x_j | x_i \rangle = \int dx_j x_j P_s(x_j |x_i)$ is the
steady state conditional average of $x_j$, $j\in {\cal N}(i)$, given $x_i$ at
site $i$.  We denote its spatial average by
\begin{equation}\label{average}
m_i= \frac{1}{N}\sum _{j \in {\cal N}(i)} \langle x_j | x_i\rangle.
\end{equation}

\noindent {\em Global coupling.}
In the case of global coupling, fluctuations of $m_i$
disappear in the limit $L \rightarrow \infty$. We thus may consider $m_i$ as a
parameter and obtain except for a constant factor a stationary solution of
(\ref{fokker})
\begin{equation}\label{statsol}
p_s(x_i,m_i) = |x_i|^{2 (a-D)/{\sigma^2}-1} e^ 
{-(x_i^2+2Dm_i/ x_i)/{\sigma^2}}.
\end{equation}
If this expression is normalizable, the stationary probability density 
$P_s(x_i,m_i)$ reads
\begin{equation} \label{statprob}
P_s(x_i,m_i)=
   \begin{cases} 
          1/{N(m_i)}\;p_s(x_i,m_i)& \text{for $x_i \in $ supp,}  \\
          0& \text{otherwise,} 
   \end{cases} 
\end{equation} 
where $N(m_i)=\int_{\textrm{supp}} dx\; p_s(x,m_i)$.  $P_s$
lives on a support on which  (\ref{statsol}) is normalizable, i.e. $N$ is
finite. 

For both  $D$ and $m_i$ nonzero the support of $P_s$ is such that $D m_i/ x_i
\geqslant 0$ ensuring normalizability of (\ref{statsol}).  For  $m_i=0$
normalizability requires that  the exponent of the algebraic factor in 
(\ref{statsol}) is larger than $-1$, i.e. $D<a$. For $D>a$ the solution
(\ref{statsol}) is not normalizable and we have  $P_s(x_i)=\delta(x_i)$. The
determination of $m_i$ is described below in detail. 

\begin{figure}[tbp] 
\begin{center} 
\epsfig{file=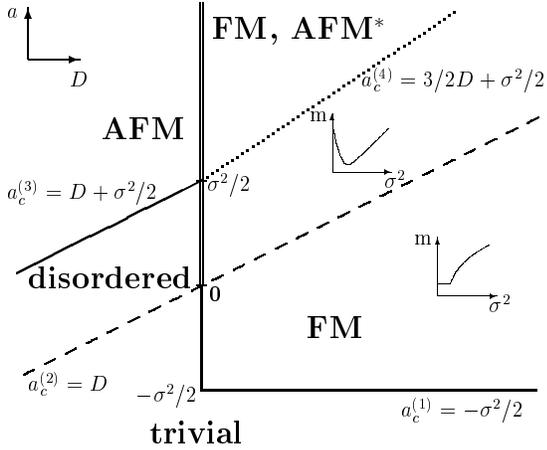,width=7.2cm}
\vspace{0.3cm} 
\caption{Phase diagram in the $(D,a)$ plane for global coupling.
Continuous transitions toward FM or AFM states occur crossing the fat solid
lines. The double solid line indicates discontinuous transitions. The critical
values $a_c^{(n)}, n=1,\dots,4$, the disordered phase, the metastable
AFM$^*$ phase, and the inserts are explained in the text.} 
\label{glphdia}
\end{center} 
\end{figure}

Varying the control parameters of the system $a$ and $D$, or the strength of
the noise $\sigma^2$, one obtains the phase diagram shown in Fig.
\ref{glphdia}.

We first consider $D>0$ which favours a FM order. In the spatially homogeneous
case $m_i\equiv m$ and for $m>0$ or $m<0$ the support of $P_s$ is $[0,\infty)$
and $(-\infty,0]$, respectively. 
All constituents have the same (statistical or
temporal) average 
\begin{equation} \label{}
\langle x\rangle=\int_{\textrm{supp}} dx \,xP_s(x,m)=\lim_{T \rightarrow \infty}
\frac{1}{T} \int_0^T dt x(t) = F(m),
\end{equation}
which equals the spatial average $m$
(ergodicity). This leads to the self-consistency condition 
\begin{equation}\label{scefm}
m=\langle x\rangle =  F(m) 
\end{equation}
determining $m$. One easily finds
\begin{equation}
F(\pm 0)=\begin{cases} \pm \sigma\Gamma(\frac{a-D}{\sigma^2}+\frac{1}{2})/
                        \Gamma(\frac{a-D}{\sigma^2})
          & \text{for $a>D$,}  \\
          0& \text{for $a<D$.} 
   \end{cases} 
\end{equation}

\begin{figure}[t]
 \begin{center}
 \epsfig{file=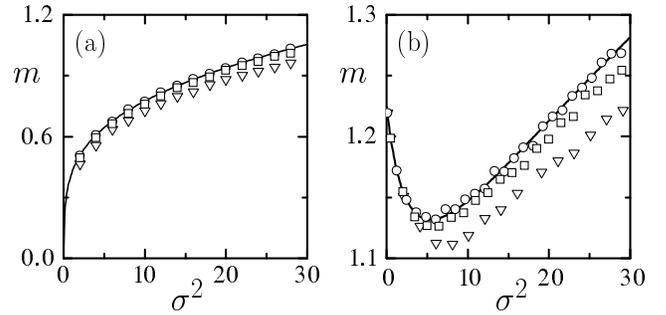,width=8.5cm}
 \caption{Order parameter $m$ vs. $\sigma^2$. 
 The solid line is the solution of (\ref{scefm}) compared with simulations of
 (\ref{langevin}) for global (circles) and n.n. coupling in $d=1$ (triangels)
 and $d=3$ (squares).  For $a<a_c^{(2)}=D$ (a) the order parameter $m$ increases
 monotonously with $\sigma^2$  whereas for $a>D$ (b) a
 pronounced minimum appears. The parameters used are 
 $a=0$ in (a) and $a=1.5$ in (b), $D=0.5$.}
 \label{m_noise} 
 \end{center}
\end{figure}

For $a<a_c^{(1)}=-\sigma^2/2$ Eq. (\ref{scefm}) has a trivial stable solution
$m=0$  which looses its stability at $a=a_c^{(1)}$ which is determined by the
condition $F'(0)=1$. It  bifurcates  into a pair of stable solutions $m=m_+>0$
and $m=-m_+=m_-$ corresponding to a continuous  transition from a paramagnetic
to a FM situation. Choosing   $m=m_+$ for instance, the stationary probability
distribution of the corresponding ergodic component is $P_s(x,m_+)$, cf. Eq.
(\ref{statprob}).   In the FM region,  for $a_c^{(1)}<a<a_c^{(2)}=D$ the
magnetization $m=\langle x\rangle$ increases monotonously with $\sigma^2$,
whereas for $a>a_c^{(2)}$ there is a nonmonotonous behaviour, cf. Fig.
\ref{m_noise}. 

\begin{figure}[b]
 \begin{center}
 \epsfig{file=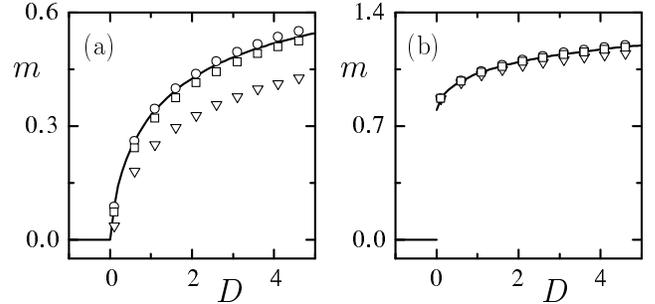,width=8.5cm}
  \caption{Order parameter $m$ vs. $D$. 
  Lines and symbols have the same meaning as in Fig. \ref{m_noise}. For
 $-\sigma^2/2=a_c^{(1)}<a<0$  the transition is continuous (a), whereas for 
 $0<a$ it is discontinuous (b). The parameters used are $a=-0.5$ in (a) and
 $a=1$ in (b), $\sigma^2=2$.
 }
 \label{m_D}
 \end{center}
\end{figure}

As a function of $D$, the  magnetization $m$ increases continuously from
zero when increasing $D$ from zero for  $a_c^{(1)}<a<0$, whereas the transition
is discontinuous for $a>0$ as shown in Fig. \ref{m_D}, cf. also
\cite{Genovese99}.

Within the FM region, a metastable \cite{footnote1} antiferromagnetic solution
(AFM$^*$) exists besides the stable FM solution for $a>a_c^{(4)}$, see Fig.
\ref{glphdia}. The critical value $a_c^{(4)}=3/2D+\sigma^2/2$ is obtained for
weak noise from an extremal approximation for $m=\langle x\rangle$ in the
spirit of \cite{Broeck97}.

For $D<0$ the situation is different. For $a<a_c^{(2)}$ we have
$P_s(x)=\delta(x)$. In the range $a_c^{(2)}<a<a_c^{(3)}=D+\sigma^2/2$ one finds
$m=0$ and the stationary probability density $P_s(x)$ lives on $(-\infty,
\infty)$, we call this the disordered phase. For $a>a_c^{(3)}$ the stationary
solution (\ref{statsol}) is normalizable only for $m\neq 0$, for $m>0$ or $m<0$ the
support is $(-\infty,0]$ or $[0,\infty)$, respectively. We define two
subsystems labelled by $+$ and $-$ for which the averages  $\langle x_i\rangle$
have $+$ or $-$ sign, respectively. For global coupling AFM
order implies $m_{\pm}\rightarrow \mp 0$ in  the limit $L \rightarrow \infty$.
Therefore, the mean values $\langle x_i\rangle=\langle x_{\pm}\rangle$ are  given by 
\begin{equation}
\label{afmgc} \langle x_{\pm}\rangle = -\langle x_\mp \rangle=   \int_0^{\pm
\infty} dx \,xP_s(x,\mp 0)=\pm F(\pm 0),   
\end{equation} 
where $P_s$ is taken from (\ref{statprob}). 

\begin{figure}[b]
 \begin{center}
 \includegraphics[width=6cm]{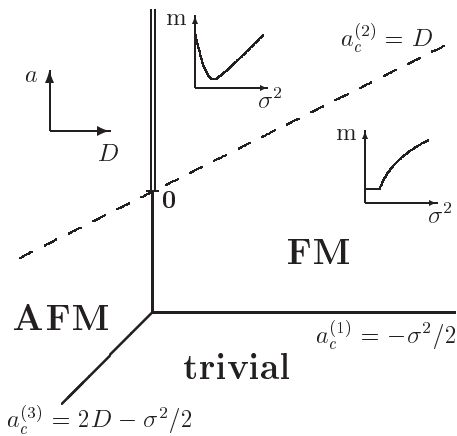}
 \vspace{0.3cm}
 \caption{Phase diagram in the $(D,a)$ plane for n.n. coupling. For $D>0$ the
 situation is the same as for global coupling except that the metastable AFM$^*$
  phase is absent here. For the case $D<0$ see text.}
 \label{nnphdia}
 \end{center}
\end{figure}

\noindent {\em Nearest neighbour coupling.}
For n.n. coupling on a cubic lattice a mean field
approximation is obtained in a similar way replacing the spatial average over
the $2d$ nearest neighbours as $m_i=1/(2d)\sum _{j \in {\cal N}(i)} \langle x_j |
x_i \rangle \approx \langle x_i \rangle$. The FM case, $D>0$, is
formally the same as for global coupling but Eq. (\ref{scefm}) holds only
approximately.
In the AFM case, $D<0$, one should take into account that now the two subsystems
$+$ and $-$ correspond to different N\'eel sublattices $A$ and $B$,
respectively, and all the nearest neighbours of a given lattice site belong to
the complementary sublattice. Self-consistency requires
\begin{equation} \label{afmnn}
m_\pm = -\langle x_\pm\rangle  =  -\int_0^{\pm \infty} dx \,xP_s(x,m_\pm)
= -m_\mp.  
\end{equation}
For $D<0$ system (\ref{langevin}) is invariant under the transformation
\cite{Birner99} $x_i\rightarrow -x_i$ for $i \in A$,  $x_j\rightarrow  x_j$ for
$j \in B$, $D\rightarrow -D$,  $a\rightarrow  a-2D$.  This implies that 
properties of the AFM phase for spatial coupling $D=-D'<0$ can be inferred from
properties of the FM phase for spatial coupling strength $D'$, cf.
\cite{Vives97}. For instance, $a_c^{(1)}=-\sigma^2/2$ transforms into
$a_c^{(3)}=2D-\sigma^2/2$. The phase diagram for n.n. coupling is shown in Fig.
\ref{nnphdia}.

\noindent
{\em Critical behaviour.}
Varying the control parameters $a$ and $\sigma^2$ one observes continuous
transitions from zero to nonzero values of $m$ with a characteristic power law
behaviour near the critical values of the control parameters. To analyze the
critical behaviour it is useful to write the self-consistency equations
(\ref{scefm}) and (\ref{afmnn}) in compact form as 
\begin{equation}\label{sceshort}
m = - \frac{2|D|}{\sigma^2} \bigg ( \frac {\partial \ln I(m)}{\partial m} \bigg 
)^{-1}\; ,
\end{equation}
where
\begin{equation}\label{int}
I(m) =  \int_0^{\infty} dx\,  x^{2 (a-D)/{\sigma^2}} e^ 
{-(x^2+2|D|m/ x)/{\sigma^2}}.
\end{equation}
In the limits $\sigma \rightarrow 0$ or $D\rightarrow\infty$ this integral can
be evaluated by the Laplace method, cf. e.g. \cite{Olver74}. Inserting the
results in (\ref{sceshort}) for small $m$, one obtains the power laws  $m
\sim (a+ \sigma^2/2+|D| - D)^{1/2}$ for $\sigma \rightarrow 0$ and $m \sim
(a+\sigma^2/2)^{1/2}$ for $D\rightarrow \infty$ with the critical exponent
$\beta =1/2$, cf. also \cite{Broeck94,Garcia96,Genovese99}.

For finite values of $\sigma$ and $D$ the scaling behaviour of $I(m)$  can be
evaluated for small $m$  with the result \cite{footnoteAsym}
\begin{equation}\label{scalI} 
I(m) \sim
m^{2(\varepsilon-D)/\sigma^2}(1+C_1m^{-2(\varepsilon-D)/\sigma^2}+
C_2m^2),
\end{equation}
where $\varepsilon=a-a_c$. The critical value $a_c$ is  $-\sigma^2/2$
for the FM case and $2D-\sigma^2/2$ for the AFM case.  Inserting
(\ref{scalI}) in Eq. (\ref{sceshort})  we  obtain for small $m$ and in lowest order
of $\varepsilon$ the power law   
\begin{equation}\label{scallaw}   m \sim
\varepsilon ^\beta, \quad\beta=\mathrm{sup} \{ 1/2,  \sigma^2/(2|D|)\},   
\end{equation}    
logarithmic corrections are easily computed. Obviously, the critical exponents
are the same varying $a$ or $\sigma^2$,  i.e. $\beta_a=\beta_{\sigma}=\beta$
using notations from \cite{Genovese99}. For models where the cubic nonlinearity
in (\ref{langevin}) is replaced by $x^{p+1}$, $p>0$,  in (\ref{scallaw}) the
value $1/2$  is replaced by $1/p$\cite{PB01}. 

\begin{figure}[tbp]
 \begin{center}
  \epsfig{file=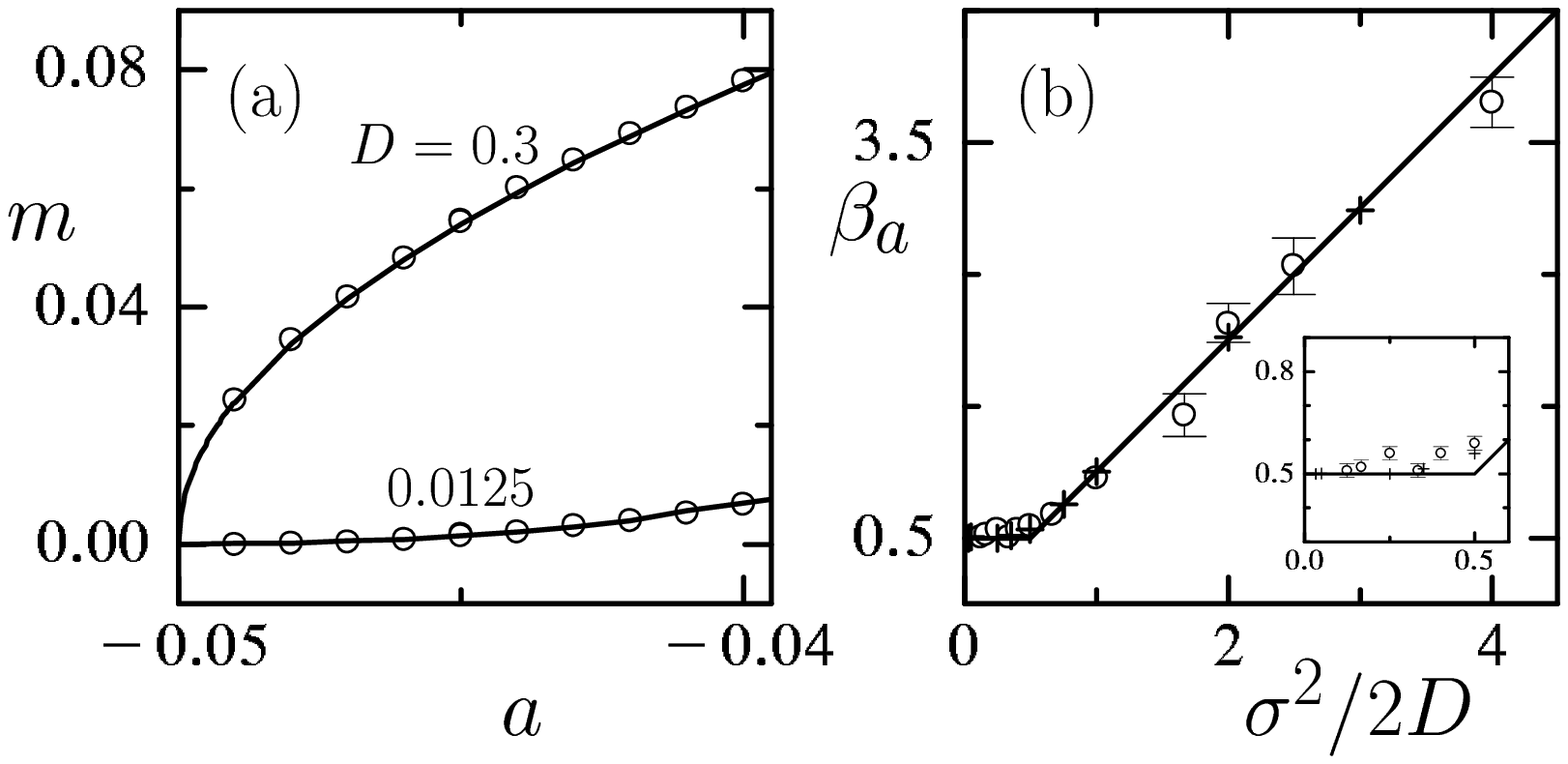,width=8.5cm}
  \epsfig{file=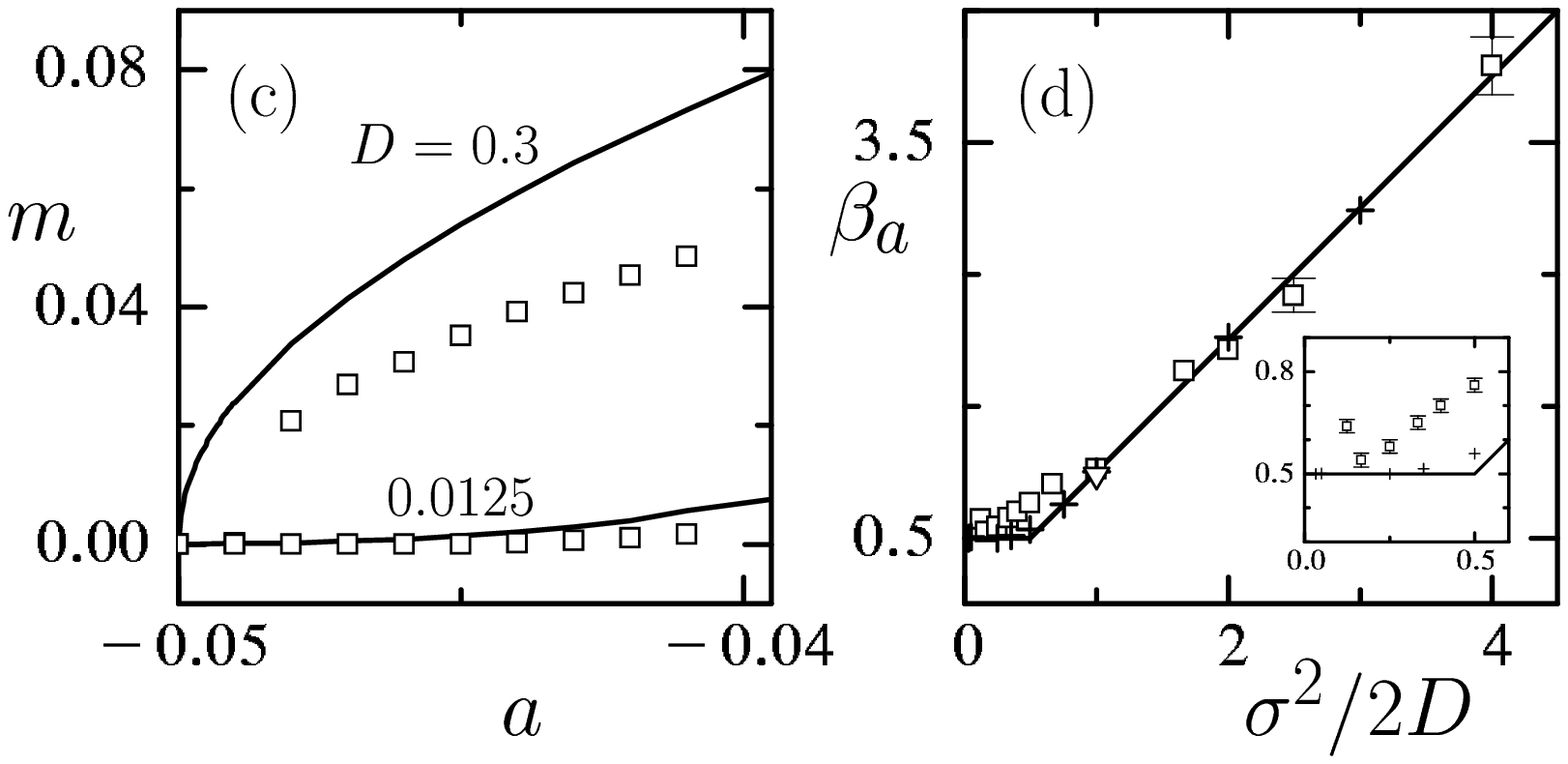,width=8.5cm}
        \caption{Critical behaviour. The Figure shows the order parameter $m$ vs.
 control parameter $a$ for different values of $D$ ($\sigma^2=0.1$, i.e.
 $a_c=-0.05$) and the  critical exponent $\beta_a$ vs. $\sigma^2/(2D)$. (a) and
 (b) refer to global coupling, (c) and (d) to n.n. coupling in $d=3$ ,
 respectively. In (a) and (c) the solid line is the numerical solution of the
 self-consistency equation (\ref{sceshort}). In (b) and (d) the solid
 line is the analytical result  (\ref{scallaw}); the $+$ symbols represent the
 numerical solution of Eq. (\ref{sceshort}). Circles and squares result from
 simulations of Eq. (\ref{langevin}); the triangle in (d) is the numerical result
 of \protect\cite{Genovese99}. Error bars are partially smaller than the symbol
 size.}
 \label{critbehav}
 \end{center}
\end{figure}

Fig. \ref{critbehav} compares the magnetization $m(a)$ and the critical
exponent $\beta$ obtained from the analytical results with simulations for
both  global and n.n. coupling \cite{footnoteNum}. For global
coupling, simulations for systems of size $10^3$ are already very close to the
results for the infinite system. 

\noindent {\em Conclusion.}  For the globally coupled model we found
analytically  a transition of the critical exponent $\beta$ from a  value
$1/2$, which reflects the order of the nonlinearity and is independent of the
strength of   noise $\sigma^2$ and spatial coupling $D$,  to a non-universal
behaviour, depending on $\sigma^2$ and $D$ independent of the order of the
nonlinearity.  This differs from the value $\beta=1$ proposed for the
continuous version of the model  in \cite{Genovese99}.

If the noise is not too strong, the 'mean field' results describe the critical
behaviour for n.n. coupling observed in our simulations very well, cf. Fig.
\ref{critbehav}d. Also the numerical result $\beta \approx 1$ obtained by
Genovese and Mu{\~n}oz \cite{Genovese99} near $a=-1$, $\sigma^2=2$ for $D=1$
(their 'weak noise phase') is in accord with our analytical result 
(\ref{scallaw}). For stronger noise, simulations for n.n. coupling may differ
considerably from the mean field prediction.

For models with a nonlinearity $x^{p+1}$, the value $1/p$ of $\beta$ is in
general different from the value $1/2$ characteristic for models with only
additive noise. In this case, one has to expect interesting crossover phenomena
for models with additive and multiplicative noise when changing the relative
strength of the noises.

This study was partially supported by the DFG with Grant BE 1417/3.

\noindent
\small 
\bibliographystyle{plain} 

\begin{thebibliography}{99}
\bibitem{footnote1}  We  prepared the system such that a fraction $\lambda$ of
the constituents have positive initial values and a fraction $1-\lambda$
negative ones.  By simulation we determined the first passage times for which
one of the constituents changes the sign. For $\lambda \neq 1/2$, the mean
first passage time (MFPT) decreases exponentially with the system size $L$ and
the system reaches the FM state very fast.  For $\lambda = 1/2$ the MFPT
increases with $L$ and is expected to diverge for $L \rightarrow \infty$.
Additive noise however naturally leads to $\lambda \neq 1/2$ and finally to a FM
state; the AFM$^*$ state is therefore only metastable, cf. also \cite{LMBKB00}.
\bibitem{footnoteAsym} Substituting  $y=x/m$ in (\ref{int}) leads to
$I(m)=m^{1+2(a-D)/\sigma^2} J(m)$. We split $J= \int_0^{y'} \ldots +
\int_{y'}^\infty \ldots = J_1 + J_2 $  at  an intermediate point $y'$ with  
$2|D|/\sigma^2 \ll y' \ll \sigma /m$. In $J_1$ and $J_2$ we expand $\exp\{-m^2
y^2/{\sigma^2}\}$ and $\exp\{-2|D|/(\sigma^2 y)\}$ to second order,
respectively. One readily obtains $J_1 \sim c_1+c_2 m^2$.    To treat $J_2$
further we  resubstitute $x=ym$ obtaining $J_2=m^{-1-2(a-D)/\sigma^2} I_2$. We
split now $I_2=  \int_{my'}^{x'}\ldots + \int_{x'}^\infty\ldots= I_2^a +I_2^b$
at a second intermediate point $x'$ with $my'<x'\ll\sigma$.  In $I_2^a$ we
expand  $\exp\{-x^2/{\sigma^2}\}$ to second order and obtain $I_2^a \sim c_3 +
c_4 m^{1+2(a-D)/{\sigma^2}} + c_5 m^{3+2(a-D)/{\sigma^2}}$; $I_2^b=c_6$. All
$c_n$,  $n=1,\dots,6$, are independent of $m$. Introducing
$\varepsilon=a-a_c$ and renaming the constants gives (\ref{scalI}).
\bibitem{footnoteNum} System (\ref{langevin}) was solved by a 4th order
Runge-Kutta scheme adopted to the stochastic case such that the noise is set
constant during one time step. The Gaussian white noise was generated by a
Box-M\"uller algorithm.  After  thermalization we computed $m$ as the temporal
average over $5 \cdot 10^4$ time steps of the spatial mean of $x_i$ over all
lattice sites. The time steps were chosen small enough to avoid numerical
instabilities ($\Delta t=10^{-3} \ldots 10^{-2}$). To determine $a_c$
numerically, we exploited that for $a<a_c$  starting with initial conditions
corresponding to $m>0$ the system relaxes in a finite time to $m'<m$. $m'$
serves as initial condition for  $a'=a+\delta a$. Iterating the procedure, $m$
relaxes further as long as we are below $a_c$, above $a_c$ it increases. The
critical exponent $\beta$ was obtained  from the slope of a linear fit of the
plot $\ln m$ vs $\ln(a-a_c)$ in the range where the first derivative of $\ln m$
is almost constant. The parameter dependence of $m$ was computed for systems of
size $L=10 \times 10 \times 10$. Critical exponents  were obtained  considering
larger systems, $L=40 \times 40 \times 40$. Periodic boundary conditions were
used for the cubic lattice.
\end{thebibliography}

\end{document}